%%%%%%%%%%%%%%%%%%%%%%%%%%%%%%%%%%%%%
%  												                    %
%                        Dan Israel and Vasilis Niarchos, 2007                                    %
%  												                    %
%%%%%%%%%%%%%%%%%%%%%%%%%%%%%%%%%%%%%
%% Draft Mode
           %turn on label
%\def\mydraft{no label}        %turn off label
   %display title on top-right

%\input lanlmac
\input harvmac

\input epsf

%\draftmode

%Macro for figure
\newcount\figno
\figno=0
\def\fig#1#2#3{
\par\begingroup\parindent=0pt\leftskip=1cm\rightskip=1cm\parindent=0pt
\baselineskip=11pt
\global\advance\figno by 1
\midinsert
\epsfxsize=#3
\centerline{\epsfbox{#2}}
\vskip 12pt
\centerline{{\bf Fig. \the\figno:~~} #1}\par
\endinsert\endgroup\par
}
\def\figlabel#1{\xdef#1{\the\figno}}

%%%%%%%
% Older defs

\def\journal#1&#2(#3){\unskip, \sl #1\ \bf #2 \rm(19#3) }
\def\andjournal#1&#2(#3){\sl #1~\bf #2 \rm (19#3) }

\def\frac#1#2{{#1\over#2}}

\def\d{\partial}

\def\inbar{\,\vrule height1.5ex width.4pt depth0pt}
\def\IC{\relax\hbox{$\inbar\kern-.3em{\rm C}$}}
\def\IR{\relax{\rm I\kern-.18em R}}
\def\IP{\relax{\rm I\kern-.18em P}}
\def\IZ{\relax{\rm I\kern-.18em Z}}
\def\IT{\relax{\rm T\kern-.60em T}}
%\def\IQ{\relax{\rm I\kern-.18em Q}}

%
%%%%%%%%%%%%%%%%%%%%%%%%%%%%%%%%%%%%
%

%
\catcode`\@=11
\def\slash#1{\mathord{\mathpalette\c@ncel{#1}}}
\overfullrule=0pt

\def\FF{{\cal F}}

\def\II{{\cal I}}

\def\KK{{\cal K}}

\def\MM{{\cal M}}
\def\NN{{\cal N}}
\def\OO{{\cal O}}
\def\PP{{\cal P}}
\def\QQ{{\cal Q}}

\def\SS{{\cal S}}
\def\TT{{\cal T}}

\def\VV{{\cal V}}

\def\ZZ{{\cal Z}}

\def\underrel#1\over#2{\mathrel{\mathop{\kern\z@#1}\limits_{#2}}}

\catcode`\@=12

%%%%%%%%%%%%%%%%%%%%%%%%%%%%%%%%%%%%%%%%%%%%%%%%%

%

\def\mod{{\rm mod}}

\def \cosh{{\rm cosh}}

\def\exp{{\rm exp}}

%%%%%%%%%%%%%%%%%%%%%%%%%%%%%%%%%%%%%%%%%%%%%%%%%
% new defs:

% Something to deal with sub-sub-sections

\def\unlockat{\catcode`\@=11}
\def\lockat{\catcode`\@=12}

\unlockat

%%% Something to deal with sub-sub-sections

\def\newsec#1{\global\advance\secno by1\message{(\the\secno. #1)}
\global\subsecno=0\global\subsubsecno=0\eqnres@t\noindent
{\bf\the\secno. #1}
\writetoca{{\secsym} {#1}}\par\nobreak\medskip\nobreak}
\global\newcount\subsecno \global\subsecno=0
\def\subsec#1{\global\advance\subsecno
by1\message{(\secsym\the\subsecno. #1)}
\ifnum\lastpenalty>9000\else\bigbreak\fi\global\subsubsecno=0
\noindent{\it\secsym\the\subsecno. #1}
\writetoca{\string\quad {\secsym\the\subsecno.} {#1}}
\par\nobreak\medskip\nobreak}
\global\newcount\subsubsecno \global\subsubsecno=0
\def\subsubsec#1{\global\advance\subsubsecno by1
\message{(\secsym\the\subsecno.\the\subsubsecno. #1)}
\ifnum\lastpenalty>9000\else\bigbreak\fi
\noindent\quad{\secsym\the\subsecno.\the\subsubsecno.}{#1}
\writetoca{\string\qquad{\secsym\the\subsecno.\the\subsubsecno.}{#1}}
\par\nobreak\medskip\nobreak}

\def\subsubseclab#1{\DefWarn#1\xdef
#1{\noexpand\hyperref{}{subsubsection}%
{\secsym\the\subsecno.\the\subsubsecno}%
{\secsym\the\subsecno.\the\subsubsecno}}%
\writedef{#1\leftbracket#1}\wrlabeL{#1=#1}}% Macros for boxes
\lockat

% End of Older Defs
%%%%%%%%%%%%%

%\def\bar{\widetilde}
\newcount\figno
\figno=0
\def\fig#1#2#3{
\par\begingroup\parindent=0pt\leftskip=1cm\rightskip=1cm\parindent=0pt
\baselineskip=11pt
\global\advance\figno by 1
\midinsert
\epsfxsize=#3
\centerline{\epsfbox{#2}}
\vskip 12pt
{\bf Fig.\ \the\figno: } #1\par
\endinsert\endgroup\par
}
\def\figlabel#1{\xdef#1{\the\figno}}
\def\encadremath#1{\vbox{\hrule\hbox{\vrule\kern8pt\vbox{\kern8pt
\hbox{$\displaystyle #1$}\kern8pt}
\kern8pt\vrule}\hrule}}
%
%

%%% Paragraphs

%%% special math symbols
\font\cmss=cmss10
\font\cmsss=cmss10 at 7pt
\def\rlx{\relax\leavevmode}
\def\inbar{\vrule height1.5ex width.4pt depth0pt}
\def\IC{\relax\,\hbox{$\inbar\kern-.3em{\rm C}$}}
\def\IN{\relax{\rm I\kern-.18em N}}
\def\IP{\relax{\rm I\kern-.18em P}}
\def\ZZ{\rlx\leavevmode\ifmmode\mathchoice{\hbox{\cmss Z\kern-.4em Z}}
 {\hbox{\cmss Z\kern-.4em Z}}{\lower.9pt\hbox{\cmsss Z\kern-.36em Z}}
 {\lower1.2pt\hbox{\cmsss Z\kern-.36em Z}}\else{\cmss Z\kern-.4em
 Z}\fi}
%%% misc.
\def\IZ{\relax\ifmmode\mathchoice
{\hbox{\cmss Z\kern-.4em Z}}{\hbox{\cmss Z\kern-.4em Z}}
{\lower.9pt\hbox{\cmsss Z\kern-.4em Z}}
{\lower1.2pt\hbox{\cmsss Z\kern-.4em Z}}\else{\cmss Z\kern-.4em
Z}\fi}
%%% misc.
\def\IZ{\relax\ifmmode\mathchoice
{\hbox{\cmss Z\kern-.4em Z}}{\hbox{\cmss Z\kern-.4em Z}}
{\lower.9pt\hbox{\cmsss Z\kern-.4em Z}}
{\lower1.2pt\hbox{\cmsss Z\kern-.4em Z}}\else{\cmss Z\kern-.4em
Z}\fi}
%%% misc.
\def\IQ{\relax\ifmmode\mathchoice
{\hbox{\cmss Q\kern-.6em Q}}{\hbox{\cmss Q\kern-.6em Q}}
{\lower.9pt\hbox{\cmsss Q\kern-.6em Q}}
{\lower1.2pt\hbox{\cmsss Q\kern-.6em Q}}\else{\cmss Q\kern-.6em
Q}\fi}

\def\narrowplus{\kern -.04truein + \kern -.03truein}
\def\narrowminus{- \kern -.04truein}
\def\narrowminussub{\kern -.02truein - \kern -.01truein}

\def\IZ{\relax\ifmmode\mathchoice
{\hbox{\cmss Z\kern-.4em Z}}{\hbox{\cmss Z\kern-.4em Z}}
{\lower.9pt\hbox{\cmsss Z\kern-.4em Z}}
{\lower1.2pt\hbox{\cmsss Z\kern-.4em Z}}\else{\cmss Z\kern-.4em
Z}\fi}
\def\IB{\relax{\rm I\kern-.18em B}}
\def\IC{{\relax\hbox{$\inbar\kern-.3em{\rm C}$}}}
\def\ID{\relax{\rm I\kern-.18em D}}
\def\IE{\relax{\rm I\kern-.18em E}}
\def\IF{\relax{\rm I\kern-.18em F}}
\def\IG{\relax\hbox{$\inbar\kern-.3em{\rm G}$}}
\def\IGa{\relax\hbox{${\rm I}\kern-.18em\Gamma$}}
\def\IH{\relax{\rm I\kern-.18em H}}
\def\II{\relax{\rm I\kern-.18em I}}
\def\IK{\relax{\rm I\kern-.18em K}}
\def\IP{\relax{\rm I\kern-.18em P}}
%\def\IX{\relax{\rm X\kern-.01em X}}
%this doesn't work

\def\mod{{\rm mod}}

\font\cmss=cmss10 \font\cmsss=cmss10 at 7pt
\def\IR{\relax{\rm I\kern-.18em R}}

\def\rrangle{{\rangle \rangle}}

%

%
%       \eqn\label{a+b=c}       gives displayed equation, numbered
%                               consecutively within sections.
%     \eqnn and \eqna define labels in advance (of eqalign?)
%
\def\eqnn#1{\xdef #1{(\secsym\the\meqno)}\writedef{#1\leftbracket#1}%
\global\advance\meqno by1\wrlabeL#1}
\def\eqna#1{\xdef #1##1{\hbox{$(\secsym\the\meqno##1)$}}
\writedef{#1\numbersign1\leftbracket#1{\numbersign1}}%
\global\advance\meqno by1\wrlabeL{#1$\{\}$}}
\def\eqn#1#2{\xdef #1{(\secsym\the\meqno)}\writedef{#1\leftbracket#1}%
\global\advance\meqno by1$$#2\eqno#1\eqlabeL#1$$}

%%%%%%%%%%%%%%%%%%%%
% References
%%%%%%%%%%%%%%%%%%%%

%\NiarchosKW
\lref\NiarchosKW{
  V.~Niarchos,
  ``Density of states and tachyons in open and closed string theory,''
  JHEP {\bf 0106}, 048 (2001)
  [arXiv:hep-th/0010154].
  %%CITATION = JHEPA,0106,048;%%
}

%\KutasovUA
\lref\KutasovUA{
  D.~Kutasov and N.~Seiberg,
  ``Noncritical superstrings,''
  Phys.\ Lett.\  B {\bf 251}, 67 (1990).
  %%CITATION = PHLTA,B251,67;%%
}

%\KutasovSV
\lref\KutasovSV{
  D.~Kutasov and N.~Seiberg,
  ``Number Of Degrees Of Freedom, Density Of States And Tachyons In String
  Theory And CFT,''
  Nucl.\ Phys.\  B {\bf 358}, 600 (1991).
  %%CITATION = NUPHA,B358,600;%%
}

%\KutasovPV
\lref\KutasovPV{
  D.~Kutasov,
  ``Some properties of (non)critical strings,''
  arXiv:hep-th/9110041.
  %%CITATION = HEP-TH/9110041;%%
}

%\FotopoulosCN
\lref\FotopoulosCN{
  A.~Fotopoulos, V.~Niarchos and N.~Prezas,
  ``D-branes and SQCD in non-critical superstring theory,''
  JHEP {\bf 0510}, 081 (2005)
  [arXiv:hep-th/0504010].
  %%CITATION = JHEPA,0510,081;%%
}

%\GomisVI
\lref\GomisVI{
  J.~Gomis and A.~Kapustin,
  ``Two-dimensional unoriented strings and matrix models,''
  JHEP {\bf 0406}, 002 (2004)
  [arXiv:hep-th/0310195].
  %%CITATION = JHEPA,0406,002;%%
}

%\BergmanYP
\lref\BergmanYP{
  O.~Bergman and S.~Hirano,
  ``The cap in the hat: Unoriented 2D strings and matrix(-vector) models,''
  JHEP {\bf 0401}, 043 (2004)
  [arXiv:hep-th/0311068].
  %%CITATION = JHEPA,0401,043;%%
}

%\AngelantonjCT
\lref\AngelantonjCT{
  C.~Angelantonj and A.~Sagnotti,
  ``Open strings,''
  Phys.\ Rept.\  {\bf 371}, 1 (2002)
  [Erratum-ibid.\  {\bf 376}, 339 (2003)]
  [arXiv:hep-th/0204089].
  %%CITATION = PRPLC,371,1;%%
}

%\BianchiYU
\lref\BianchiYU{
  M.~Bianchi and A.~Sagnotti,
  ``On the systematics of open string theories,''
  Phys.\ Lett.\  B {\bf 247}, 517 (1990).
  %%CITATION = PHLTA,B247,517;%%
}

%\SagnottiGA
\lref\SagnottiGA{
  A.~Sagnotti,
  ``Some properties of open string theories,''
  arXiv:hep-th/9509080.
  %%CITATION = HEP-TH/9509080;%%
}

%\SagnottiQJ
\lref\SagnottiQJ{
  A.~Sagnotti,
  ``Surprises in open-string perturbation theory,''
  Nucl.\ Phys.\ Proc.\ Suppl.\  {\bf 56B}, 332 (1997)
  [arXiv:hep-th/9702093].
  %%CITATION = NUPHZ,56B,332;%%
}

%\KlebanovYY
\lref\KlebanovYY{
  I.~R.~Klebanov and A.~A.~Tseytlin,
  ``D-branes and dual gauge theories in type 0 strings,''
  Nucl.\ Phys.\  B {\bf 546}, 155 (1999)
  [arXiv:hep-th/9811035].
  %%CITATION = NUPHA,B546,155;%%
}

%\WittenKH
\lref\WittenKH{
  E.~Witten,
  ``Baryons In The 1/N Expansion,''
  Nucl.\ Phys.\  B {\bf 160}, 57 (1979).
  %%CITATION = NUPHA,B160,57;%%
}

%\tHooftJZ
\lref\tHooftJZ{
  G.~'t Hooft,
  ``A planar diagram theory for strong interactions,''
  Nucl.\ Phys.\  B {\bf 72}, 461 (1974).
  %%CITATION = NUPHA,B72,461;%%
}

%\Coleman
\lref\Coleman{
S.~Coleman,
``$1/N$, in Aspects of Symmetry,"
Cambridge University Press, Cambridge, 1985.
}

%\NakayamaEP
\lref\NakayamaEP{
  Y.~Nakayama,
  ``Tadpole cancellation in unoriented Liouville theory,''
  JHEP {\bf 0311}, 017 (2003)
  [arXiv:hep-th/0309063].
  %%CITATION = JHEPA,0311,017;%%
}

%\DouglasUP
\lref\DouglasUP{
  M.~R.~Douglas, I.~R.~Klebanov, D.~Kutasov, J.~M.~Maldacena, E.~Martinec and N.~Seiberg,
  ``A new hat for the c = 1 matrix model,''
  arXiv:hep-th/0307195.
  %%CITATION = HEP-TH/0307195;%%
}

%\TakayanagiSM
\lref\TakayanagiSM{
  T.~Takayanagi and N.~Toumbas,
  ``A matrix model dual of type 0B string theory in two dimensions,''
  JHEP {\bf 0307}, 064 (2003)
  [arXiv:hep-th/0307083].
  %%CITATION = JHEPA,0307,064;%%
}

%\MurthyES
\lref\MurthyES{
  S.~Murthy,
  ``Notes on non-critical superstrings in various dimensions,''
  JHEP {\bf 0311}, 056 (2003)
  [arXiv:hep-th/0305197].
  %%CITATION = JHEPA,0311,056;%%
}

%\DudasND
\lref\DudasND{
  E.~Dudas, G.~Pradisi, M.~Nicolosi and A.~Sagnotti,
  ``On tadpoles and vacuum redefinitions in string theory,''
  Nucl.\ Phys.\  B {\bf 708}, 3 (2005)
  [arXiv:hep-th/0410101].
  %%CITATION = NUPHA,B708,3;%%
}

%\IsraelIR
\lref\IsraelIR{
  D.~Israel, C.~Kounnas, A.~Pakman and J.~Troost,
  ``The partition function of the supersymmetric two-dimensional black hole
  and little string theory,''
  JHEP {\bf 0406}, 033 (2004)
  [arXiv:hep-th/0403237].
  %%CITATION = JHEPA,0406,033;%%
}

%\AharonyUB
\lref\AharonyUB{
  O.~Aharony, M.~Berkooz, D.~Kutasov and N.~Seiberg,
  ``Linear dilatons, NS5-branes and holography,''
  JHEP {\bf 9810}, 004 (1998)
  [arXiv:hep-th/9808149].
  %%CITATION = JHEPA,9810,004;%%
}

%\GiveonZM
\lref\GiveonZM{
  A.~Giveon, D.~Kutasov and O.~Pelc,
  ``Holography for non-critical superstrings,''
  JHEP {\bf 9910}, 035 (1999)
  [arXiv:hep-th/9907178].
  %%CITATION = JHEPA,9910,035;%%
}

%\SeibergZK
\lref\SeibergZK{  N.~Seiberg,
  ``New theories in six dimensions and matrix description of M-theory on  T**5
  and T**5/Z(2),''
  Phys.\ Lett.\  B {\bf 408} (1997) 98
  [arXiv:hep-th/9705221].
  %%CITATION = PHLTA,B408,98;%%
}

%\GiveonPX
\lref\GiveonPX{
  A.~Giveon and D.~Kutasov,
  ``Little string theory in a double scaling limit,''
  JHEP {\bf 9910}, 034 (1999)
  [arXiv:hep-th/9909110].
  %%CITATION = JHEPA,9910,034;%%
}

%\MurthyXT
\lref\MurthyXT{
  S.~Murthy and J.~Troost,
  ``D-branes in non-critical superstrings and duality in N = 1 gauge theories
  with flavor,''
  JHEP {\bf 0610}, 019 (2006)
  [arXiv:hep-th/0606203].
  %%CITATION = JHEPA,0610,019;%%
}

%\IsraelSI
\lref\IsraelSI{
  D.~Israel and V.~Niarchos,
  ``Orientifolds in N=2 Liouville Theory and its Mirror,''
  arXiv:hep-th/0703151.
  %%CITATION = HEP-TH/0703151;%%
}

%\AngelantonjGJ
\lref\AngelantonjGJ{
  C.~Angelantonj,
  ``Non-tachyonic open descendants of the 0B string theory,''
  Phys.\ Lett.\  B {\bf 444}, 309 (1998)
  [arXiv:hep-th/9810214].
  %%CITATION = PHLTA,B444,309;%%
}

%\BlumenhagenNS
\lref\BlumenhagenNS{
  R.~Blumenhagen, A.~Font and D.~Lust,
  ``Tachyon-free orientifolds of type 0B strings in various dimensions,''
  Nucl.\ Phys.\  B {\bf 558}, 159 (1999)
  [arXiv:hep-th/9904069].
  %%CITATION = NUPHA,B558,159;%%
}

%\BlumenhagenAD
\lref\BlumenhagenAD{
  R.~Blumenhagen and A.~Kumar,
  ``A note on orientifolds and dualities of type 0B string theory,''
  Phys.\ Lett.\  B {\bf 464}, 46 (1999)
  [arXiv:hep-th/9906234].
  %%CITATION = PHLTA,B464,46;%%
}

%\ForgerEV
\lref\ForgerEV{
  K.~Forger,
  ``On non-tachyonic Z(N) x Z(M) orientifolds of type 0B string theory,''
  Phys.\ Lett.\  B {\bf 469}, 113 (1999)
  [arXiv:hep-th/9909010].
  %%CITATION = PHLTA,B469,113;%%
}

%\BlumenhagenUY
\lref\BlumenhagenUY{
  R.~Blumenhagen, A.~Font and D.~Lust,
  ``Non-supersymmetric gauge theories from D-branes in type 0 string  theory,''
  Nucl.\ Phys.\  B {\bf 560}, 66 (1999)
  [arXiv:hep-th/9906101].
  %%CITATION = NUPHA,B560,66;%%
}

%\AngelantonjQG
\lref\AngelantonjQG{
  C.~Angelantonj and A.~Armoni,
  ``Non-tachyonic type 0B orientifolds, non-supersymmetric gauge theories  and
  cosmological RG flow,''
  Nucl.\ Phys.\  B {\bf 578}, 239 (2000)
  [arXiv:hep-th/9912257].
  %%CITATION = NUPHA,B578,239;%%
}

%\ItzhakiZR
\lref\ItzhakiZR{
  N.~Itzhaki, D.~Kutasov and N.~Seiberg,
  ``Non-supersymmetric deformations of non-critical superstrings,''
  JHEP {\bf 0512}, 035 (2005)
  [arXiv:hep-th/0510087].
  %%CITATION = JHEPA,0512,035;%%
}

%\HarmarkSF
\lref\HarmarkSF{
  T.~Harmark, V.~Niarchos and N.~A.~Obers,
  ``Stable non-supersymmetric vacua in the moduli space of non-critical
  superstrings,''
  Nucl.\ Phys.\  B {\bf 759}, 20 (2006)
  [arXiv:hep-th/0605192].
  %%CITATION = NUPHA,B759,20;%%
}

%\KiritsisTA
\lref\KiritsisTA{
E.~Kiritsis and C.~Kounnas,
  ``Infrared Regularization Of Superstring Theory And The One Loop Calculation
  Of Coupling Constants,''
  Nucl.\ Phys.\  B {\bf 442} (1995) 472
  [arXiv:hep-th/9501020].
  %%CITATION = NUPHA,B442,472;%%
}

%\DienesES
\lref\DienesES{
  K.~R.~Dienes,
  ``Modular invariance, finiteness, and misaligned supersymmetry: New
  constraints on the numbers of physical string states,''
  Nucl.\ Phys.\  B {\bf 429}, 533 (1994)
  [arXiv:hep-th/9402006].
  %%CITATION = NUPHA,B429,533;%%
}

%\DienesPM
\lref\DienesPM{
  K.~R.~Dienes, M.~Moshe and R.~C.~Myers,
  ``String Theory, Misaligned Supersymmetry, And The Supertrace Constraints,''
  Phys.\ Rev.\ Lett.\  {\bf 74}, 4767 (1995)
  [arXiv:hep-th/9503055].
  %%CITATION = PRLTA,74,4767;%%
}

%\GaberdielJR
\lref\GaberdielJR{
  M.~R.~Gaberdiel,
  ``Lectures on non-BPS Dirichlet branes,''
  Class.\ Quant.\ Grav.\  {\bf 17}, 3483 (2000)
  [arXiv:hep-th/0005029].
  %%CITATION = CQGRD,17,3483;%%
}

%\BrunnerEM
\lref\BrunnerEM{
  I.~Brunner and K.~Hori,
  ``Notes on orientifolds of rational conformal field theories,''
  JHEP {\bf 0407}, 023 (2004)
  [arXiv:hep-th/0208141].
  %%CITATION = JHEPA,0407,023;%%
}

%\BrunnerZM
\lref\BrunnerZM{
  I.~Brunner and K.~Hori,
  ``Orientifolds and mirror symmetry,''
  JHEP {\bf 0411}, 005 (2004)
  [arXiv:hep-th/0303135].
  %%CITATION = JHEPA,0411,005;%%
}

%\MurthyQM
\lref\MurthyQM{
  S.~Murthy,
  ``On supersymmetry breaking in string theory from gauge theory in a throat,''
  arXiv:hep-th/0703237.
  %%CITATION = HEP-TH/0703237;%%
}

%\AshokSF
\lref\AshokSF{
  S.~K.~Ashok, S.~Murthy and J.~Troost,
  ``D-branes in unoriented non-critical strings and duality in SO(N) and Sp(N)
  gauge theories,''
  arXiv:hep-th/0703148.
  %%CITATION = HEP-TH/0703148;%%
}

%\AshokPY
\lref\AshokPY{
  S.~K.~Ashok, S.~Murthy and J.~Troost,
  ``D-branes in non-critical superstrings and minimal super Yang-Mills in
  various dimensions,''
  Nucl.\ Phys.\  B {\bf 749}, 172 (2006)
  [arXiv:hep-th/0504079].
  %%CITATION = NUPHA,B749,172;%%
}

%\FischlerCI
\lref\FischlerCI{
  W.~Fischler and L.~Susskind,
  ``Dilaton Tadpoles, String Condensates And Scale Invariance,''
  Phys.\ Lett.\  B {\bf 171}, 383 (1986).
  %%CITATION = PHLTA,B171,383;%%
}

%\FischlerTB
\lref\FischlerTB{
  W.~Fischler and L.~Susskind,
  ``Dilaton Tadpoles, String Condensates And Scale Invariance. 2,''
  Phys.\ Lett.\  B {\bf 173}, 262 (1986).
  %%CITATION = PHLTA,B173,262;%%
}

%\DasDY
\lref\DasDY{
  S.~R.~Das and S.~J.~Rey,
  ``Dilaton Condensates and Loop Effects in Open and Closed Bosonic Strings,''
  Phys.\ Lett.\  B {\bf 186}, 328 (1987).
  %%CITATION = PHLTA,B186,328;%%
}

%\CallanAT
\lref\CallanAT{
  C.~G.~Callan, J.~A.~Harvey and A.~Strominger,
  ``Supersymmetric string solitons,''
  arXiv:hep-th/9112030.
  %%CITATION = HEP-TH/9112030;%%
}

%\DiVecchiaEV
\lref\DiVecchiaEV{
  P.~Di Vecchia, A.~Liccardo, R.~Marotta and F.~Pezzella,
  ``Brane-inspired orientifold field theories,''
  JHEP {\bf 0409}, 050 (2004)
  [arXiv:hep-th/0407038].
  %%CITATION = JHEPA,0409,050;%%
}

%\DudasFF
\lref\DudasFF{
  E.~Dudas and J.~Mourad,
  ``Brane solutions in strings with broken supersymmetry and dilaton
  tadpoles,''
  Phys.\ Lett.\  B {\bf 486}, 172 (2000)
  [arXiv:hep-th/0004165].
  %%CITATION = PHLTA,B486,172;%%
}

%\BlumenhagenDC
\lref\BlumenhagenDC{
  R.~Blumenhagen and A.~Font,
  ``Dilaton tadpoles, warped geometries and large extra dimensions for
  non-supersymmetric strings,''
  Nucl.\ Phys.\  B {\bf 599}, 241 (2001)
  [arXiv:hep-th/0011269].
  %%CITATION = NUPHA,B599,241;%%
}

%\HosomichiPH
\lref\HosomichiPH{
  K.~Hosomichi,
  ``N = 2 Liouville theory with boundary,''
  JHEP {\bf 0612}, 061 (2006)
  [arXiv:hep-th/0408172].
  %%CITATION = JHEPA,0612,061;%%
}

%\BergmanRF
\lref\BergmanRF{
  O.~Bergman and M.~R.~Gaberdiel,
  ``A non-supersymmetric open-string theory and S-duality,''
  Nucl.\ Phys.\  B {\bf 499}, 183 (1997)
  [arXiv:hep-th/9701137].
  %%CITATION = NUPHA,B499,183;%%
}

%\BachasID
\lref\BachasID{
  C.~Bachas, N.~Couchoud and P.~Windey,
  ``Orientifolds of the 3-sphere,''
  JHEP {\bf 0112} (2001) 003
  [arXiv:hep-th/0111002].
%%CITATION = JHEPA,0112,003;%%
}

%\BanksMN
\lref\BanksMN{
  T.~Banks,
  ``Matrix theory,''
  Nucl.\ Phys.\ Proc.\ Suppl.\  {\bf 67}, 180 (1998)
  [arXiv:hep-th/9710231].
  %%CITATION = NUPHZ,67,180;%%
}

%\AngelantonjTS
\lref\AngelantonjTS{
  C.~Angelantonj and E.~Dudas,
  ``Metastable String Vacua,''
  arXiv:0704.2553 [hep-th].
  %%CITATION = ARXIV:0704.2553;%%
}

%\ForsteUR
\lref\ForsteUR{
  S.~Forste, D.~Ghoshal and S.~Panda,
  ``An orientifold of the solitonic fivebrane,''
  Phys.\ Lett.\  B {\bf 411} (1997) 46
  [arXiv:hep-th/9706057].
  %%CITATION = PHLTA,B411,46;%%
}

%BurshtynXZ
\lref\BurshtynXZ{
  D.~Burshtyn, S.~Elitzur and Y.~Mandelbaum,
  ``Probing orientifold behavior near NS branes,''
  JHEP {\bf 0403} (2004) 022
  [arXiv:hep-th/0311188].
  %%CITATION = JHEPA,0403,022;%%
}

%%%%%%%%%%%%%%%%%%%%%
%                      Title Page                              %
%%%%%%%%%%%%%%%%%%%%%
\Title{
\vbox{\hbox{CPHT-RR021.0407}}
}
{\vbox{
\centerline{Tree-Level Stability Without Spacetime Fermions:}
\vskip0.4cm
\centerline{Novel Examples in String Theory}}}

\vskip -.5cm

\centerline{Dan Isra\"el$^\dagger$ and
Vasilis Niarchos$^\ddagger$\footnote{$^\diamond$}{israel@iap.fr, 
niarchos@cpht.polytechnique.fr}}

\vskip .2in

\centerline{\it $^\dagger$GRECO, Institut d'Astrophysique de Paris}
\centerline{\it 98bis Bd Arago, 75014 Paris, France\foot{Unit\'e 
mixte de Recherche 7095, CNRS -- Universit\'e Pierre et Marie Curie}}
\vskip .4cm
\centerline{\it $^\ddagger$Centre de Physique Th\'eorique, \'Ecole Polytechnique}
\centerline{\it 91128 Palaiseau, France\foot{Unit\'e mixte de Recherche
7644, CNRS -- \'Ecole Polytechnique}}

\vskip .8cm

%%abstract
\noindent
Is perturbative stability intimately tied with the existence of spacetime
fermions in string theory in more than two dimensions?
Type 0$'$B string theory in ten-dimensional flat space is a rare example
of a non-tachyonic, non-supersymmetric string theory with a
purely bosonic closed string spectrum. However, all
known type 0$'$ constructions exhibit massless NSNS tadpoles
signaling the fact that we are not expanding around a true vacuum
of the theory. In this note, we are searching for perturbatively stable examples
of type 0$'$ string theory without massless tadpoles in backgrounds with a spatially
varying dilaton. We present two examples
with this property in non-critical string theories that exhibit
four- and six-dimensional Poincar\'e invariance. We discuss the D-branes
that can be embedded in this context and the type of gauge theories
that can be constructed in this manner. We also comment on the
embedding of these non-critical models in critical string theories and
their holographic (Little String Theory) interpretation
and propose a general conjecture for the role of asymptotic
supersymmetry in perturbative string theory.

\vskip .2cm

\Date{May, 2007}

\vfill

\listtoc
\writetoc

\newsec{Introduction and summary}

Supersymmetry has been a major driving force in
theoretical physics for the last two or three decades. It is not
only a conceptually appealing construction and a powerful
calculational tool, but is most likely also an important ingredient
of the real world. At the same time, there is an obvious interest
in situations where supersymmetry has been broken mildly or is
altogether absent. In this note, we will be interested in
string theory examples where the spectrum of closed strings
is purely bosonic, so supersymmetry in the bulk is broken to the 
highest degree. Weakly coupled string theories with this feature are 
expected to be relevant for the holographic description of certain
non-supersymmetric large $N$ gauge theories, for instance large $N$ QCD.

A well known example of a string theory with a purely bosonic
closed string spectrum is type 0 string theory, but in
this case the spectrum contains a closed string tachyon signaling the
perturbative instability of the vacuum. The type 0 closed string tachyon 
can be projected out of the physical spectrum in a non-oriented version
of this theory, which is known as the type 0$'$B theory 
\refs{\BianchiYU,\SagnottiGA,\SagnottiQJ}. However, typically in this case there 
are massless NSNS tadpoles at tree-level which
drive the theory towards the true vacuum via the Fischler-Susskind
mechanism \refs{\FischlerCI,\FischlerTB,\DasDY}. 
In this new vacuum, which may or may not be perturbative, 
the shifted background is expected to have a non-trivial metric and dilaton.
In fact, in cases where the backreaction of the NSNS tadpoles has been
analyzed \refs{\DudasFF,\BlumenhagenDC} (see also \AngelantonjTS\
for a recent discussion) it was found that the system
is naturally driven at strong coupling. Furthermore, 
at higher genus, one obtains additional tadpoles for the dilaton as
an immediate consequence of the absence of supersymmetry. 

As a simple attempt to get rid of the massless tadpoles,
we consider in section 2 type 0B string theory compactified on $\IR^{8,1}\times S^1$ with
a modified worldsheet parity projection that acts non-trivially on the $S^1$. This example
involves a space-filling O$'$9 orientifold, constructed in
such a way that it sources only massive RR fields. Although
the ground state of the NSNS sector (with zero momentum
and winding along the $S^1$) is projected out, a subset
of more massive states with non-vanishing momentum and/or winding remains.
It turns out that for any value of the compatification radius there are
tachyonic modes that cannot be projected out.
This paradigm demonstates how non-trivial it is, in general,
to obtain an example in type 0$'$ string theory that is at the same time
non-tachyonic and does not exhibit massless tadpoles, $i.e.$ that exhibits
full tree-level stability. 

The generic presence of dilaton tadpoles in the above examples motivates
the study of type 0$'$ string theories in backgrounds with a non-trivial dilaton profile.
In section~3, we present the main result of this note, namely type 0$'$ string theories in
non-critical dimensions, with an asymptotic linear dilaton, that are perturbatively stable.
Similar to the ten-dimensional model discussed in section 2, these are also situations with
space-filling orientifolds sourcing only massive RR fields. However,
the spectrum in this case is non-tachyonic. The orientifold projection 
removes the closed string tachyons from the continuous representations
(closed strings that propagate along the linear dilaton direction). 
The models we consider have no localized tachyons, 
so the whole closed string spectrum is tachyon-free.
The origin of this crucial difference with the standard examples
of type 0$'$ lies in the presence of the
background charge of the linear dilaton, that gives a universal
positive mass shift to the spectrum, and the non-trivial diagonal
GSO projection in non-critical superstrings (for the six dimensional example). 

The vacua presented in section 3 are, to our knowledge, the
first $bona$ $fide$ examples of weakly coupled string theories
with a purely bosonic closed string spectrum and full tree-level stability
($i.e.$ are tachyon- and tadpole-free).
Are these examples special isolated cases, or part of a more general
construction with the same features? The possibility of more
general type 0$'$ vacua with a linear dilaton is discussed in
the concluding section. To address this issue we discuss
the embedding of these vacua in ten-dimensional string theory
and their relation with fivebrane configurations
and Little String Theory. We find that our non-critical type 0$'$ vacua
arise as an equivalent description of the near-horizon limit of NS5-branes configurations
in ten-dimensional type 0 string theories in the presence of orientifold planes. For instance,
the six-dimensional model corresponds to the double scaling limit of a pair of 
parallel fivebranes in type 0A with an O$'$6-plane at an equal distance between them.

This correspondence between non-critical type 0$'$B and fivebranes/orientifold 
configurations leads to a number of interesting observations. First, in the near-horizon 
geometry of more than two parallel fivebranes, the O$'$6-plane
does not succeed in fully projecting out the tachyon. 
Perturbative stability in these models is only possible deep in the stringy
regime, for a pair of fivebranes. Secondly, the non-critical models at hand 
correspond to fivebranes/orientifold configurations immersed in a type 0A bulk 
that contains a closed string tachyon.  
In the near-horizon decoupling limit of two fivebranes only an
$s$-wave mode of the tachyon remains and this is projected out
by the orientifold.

In the last part of the concluding
section, we elaborate on the general role of spacetime fermions and supersymmetry 
in perturbatively stable string theory and propose a new conjecture 
that links perturbative stability and asymptotic supersymmetry generalizing
previous arguments in \refs{\KutasovSV\DienesES\DienesPM-\NiarchosKW}
in accordance with the specific examples presented in this note.

Another motivation behind this work is holography for non-supersymmetric
gauge theories. For instance, by placing $N$ D3-branes on top of an O$'$3-plane
in ten-dimensional flat spacetime, one obtains on the D-branes a 
non-supersymmetric gauge theory with $U(N)$ gauge group.
The AdS/CFT correspondence for this theory, which is similar
in spirit with the type 0 AdS/CFT proposal of \KlebanovYY, was discussed
in \AngelantonjQG. Further examples of this sort, in the presence or absence
of bulk tachyons, can be found in \refs{\BlumenhagenUY,\AngelantonjQG,\DiVecchiaEV}
and references therein.  D-branes in the type IIA variant of  the non-critical string theories
considered in this note can be used to engineer interesting four dimensional gauge
theories like $\NN=1$ SQCD \FotopoulosCN. Investigating the D-branes
of the four-dimensional non-critical type 0$'$ string theory of section 3 we find that the
corresponding D-branes are unstable and give rise to four-dimensional non-supersymmetric
gauge theories in the presence of an open string tachyon. Nevertheless,
it is possible in this context to engineer odd dimensional (flavored)
non-supersymmetric gauge theories by using D-branes of a different
dimensionality. Non-supersymmetric four-dimensional gauge theories
can be realized on D-branes without open string tachyons in the six-dimensional 
type 0$'$ model. We comment on the various possibilities in section 3.

\newsec{Critical type $0'$B strings}

\subsec{Overview}

Tachyon-free type 0 models in ten-dimensional critical string theory
were first discussed in \refs{\BianchiYU,\SagnottiGA,\SagnottiQJ}.
In the original example, the worldsheet theory of type 0B strings
is modded out by the modified worldsheet parity $\Omega'=(-)^{\bar F} \Omega$,
where $\bar F$ denotes the right-moving worldsheet fermion number.
This parity acts in the closed string sector as\foot{By definition $\Omega$
acts on fermion bilinears as $\psi_r \bar \psi_r \to \psi_r \bar\psi_r$ and
$\Omega'$ as $\psi_r \bar \psi_r \to -\psi_r \bar \psi_r$.}
\eqn\aaa{
\Omega' \alpha_n \Omega'=\bar \alpha_n~, ~ ~
\Omega' \psi_r \Omega'=\bar \psi_r~,~ ~ 
\Omega' \bar \psi_r \Omega'=\psi_r~, ~ ~
\Omega' |0\rangle_{NSNS}=-|0\rangle_{NSNS}
}
and projects out the tachyonic ground state in the NSNS sector,
which is part of the type 0B spectrum. 
In spacetime, this action introduces a space-filling O$'$9-plane, which is 
a source for the non-dynamical RR ten-form. 
Consistency requires the cancellation of this tadpole. This can be achieved
with the addition of thirty two D9- and D9$'$-branes (for a detailed 
description of D-branes in type 0B string theory see \GaberdielJR).
However, the addition of D-branes leads to an uncancelled dilaton
tadpole from the NSNS sector. This tadpole does not render the theory
inconsistent, but shows that we are not expanding around a true vacuum
of the theory. Since it is hard to determine the true vacuum ($e.g.$ by implementing
the Fischler-Susskind mechanism) and because the true vacuum 
may be naturally at strong coupling, it is unclear to what extend we can read
off the true properties of the theory from the above construction
(see however \DudasND\ for a more recent discussion of tadpoles and vacuum
redefinitions in string theory). 

Despite this annoying feature, the above construction leads
to a very interesting non-supersymmetric model. In the bulk, this is a purely
bosonic theory of unoriented closed strings with a massless spectrum, which
is essentially a bosonic truncation of the type IIB spectrum. The only difference
is the absence of the NSNS  two-form field, which is projected
out by $\Omega'$. On the D9-, D9$'$-branes one obtains a bosonic spectrum 
from open strings stretching between the same type of branes and a 
fermionic spectrum from open strings stretching between a D9 and a D9$'$. 
At the massless level, the former give a $U(32)$ gauge theory and the latter 
a Majorana-Weyl fermion in the ${\bf 496}\oplus\overline{\bf 496}$ of $U(32)$.

An exhaustive analysis of tachyon-free orbifold compactifications 
has been performed in \refs{\AngelantonjGJ\BlumenhagenNS\BlumenhagenAD-\ForgerEV}
(for a review see \AngelantonjCT). Typically in these examples, one finds 
twisted sector closed string tachyons that cannot be projected out
by any generalized $\Omega$ projection. There are certain cases, however,
where twisted sector tachyons are absent and non-tachyonic vacua can
be constructed \AngelantonjCT.

The presence of massless NSNS and RR tadpoles is a generic feature
in these examples. Canceling the RR tadpoles requires the addition
of D-branes and one is usually left with a theory of open and closed strings
in the presence of an uncanceled dilaton tadpole. Are massless tadpoles at tree level
an essential property of all type 0$'$ vacua? Is it always necessary to add
a certain number of D-branes? The answer to both of these questions is no. 
In the next section, we present two examples of non-critical type 0$'$ closed strings 
in more than two dimensions, where no tachyons and no massless tadpoles exist and no 
D-branes need to be added for consistency. We are not aware of a similar example
on backgrounds with a constant dilaton. To illustrate the rarity of such theories,
we present in the next subsection a $0'$ orientifold of type 0B theory on
$\IR^{8,1}\times S^1$, which does not exhibit any massless tadpoles,
but fails to fully project out the zero modes of the closed string tachyon.

\subsec{A model without massless tadpoles but with tachyons}

Let us consider type 0B string theory on $\IR^{8,1} \times S^1$.\foot{The
example in this subsection was suggested to us by Carlo Angelantonj.}
This theory has the torus partition function
\eqn\aab{
\TT=\frac{\VV_{10}}{2}\int_{\FF} \frac{d\tau d\bar \tau}{4\tau_2}
\frac{1}{(8\pi^2 \tau_2)^{9/2}}\frac{1}{\eta^8\bar \eta^8}
\sum_{n,w=-\infty}^\infty q^{\frac{1}{2}(\frac{n}{R}+\frac{wR}{2})^2}
\bar q^{\frac{1}{2}(\frac{n}{R}-\frac{wR}{2})^2}
\sum_{a,b \in \IZ_2} \frac{\vartheta^4 \big[ {a \atop b}\big]\bar 
\vartheta^4\big[ {a \atop b} \big]}
{\eta^4 \bar \eta^4}
}
where $R$ denotes the $S^1$ radius, $\VV_{10}$  the overall volume diverging factor, 
$\tau$  the modular parameter of the torus, 
$q=e^{2\pi i \tau}$ and $\FF$ is the $PSL(2,\IZ)$ fundamental domain.
The modular functions $\vartheta \big[ {a \atop b} \big](\tau)$, $\eta(\tau)$ 
are the standard $\vartheta$- and $\eta$-functions.

As an attempt to obtain a model similar to type 
0$'$B but with a massive RR tadpole, we want to mod out this
theory with the worldsheet parity
\eqn\aac{
\PP=s(-)^{\bar F} \Omega
~,}
where $s$ is a shift symmetry acting on $X$, the  $U(1)$ boson of radius $R$, 
that parametrizes the $S^1$ part of the background:
\eqn\aad{
s~: ~ X \to X+\pi R
~.}
In the NSNS sector, $\PP$
acts on the states $|n,w\rangle_{NSNS}$ (with momentum $n$ and 
winding $w$ around the $S^1$) as
\eqn\aae{
\PP~:~|n,w\rangle \to (-)^{n+1} |n,-w\rangle
~.}
In particular, $\PP$ projects out all the NSNS ground states with
even momentum and zero winding.
The lightest allowed modes in the NSNS sector are 
($i$) the antisymmetric combination of the 
 modes with zero momentum and winding number $w=\pm 1$ and
($ii$) states with momentum $n=\pm 1$ and zero winding.
The winding (respectively momentum) modes have mass squared 
(in our conventions $\alpha'=2$)
\eqn\aaea{
m_{w=\pm 1}^2=-1+\frac{R^2}{4}~, ~ ~ 
m_{n=\pm 1}^2=-1+\frac{1}{R^2}
~.} 
It is clear from these expressions that there is no value of the radius
$R$ where both $m^2_{w=\pm 1}$ and $m^2_{n=\pm 1}$ are simultaneously
positive. Hence, the bulk tachyon cannot be projected out
fully in this example. 

In the Klein bottle amplitude only states with zero 
winding contribute. In the direct channel one obtains
\eqn\aaf{
\KK=\frac{\VV_{10}}{2} \int_0^\infty \frac{dt}{2t}~ 
\frac{1}{(8\pi^2 t)^{9/2}} \frac{1}{\eta^8(2it)}
\sum_{n \in \IZ} (-)^{n+1} e^{-2\pi t \frac{n^2}{R^2}}
\sum_{a\in \IZ_2} (-)^a \frac{\vartheta^4\big[ {a \atop 1} \big](2it)}
{\eta^4(2it)}
~,}
where only the $\widetilde{\rm NS}$$\widetilde{\rm NS}$ and
$\widetilde{\rm R}$$\widetilde{\rm R}$ sectors appear.
Using the well-known modular transformation properties of the 
$\vartheta$- and $\eta$-functions and the Poisson resummation
formula we can easily deduce the expression of the Klein bottle amplitude
in the transverse crosscap channel ($l=\frac{1}{2t}$)
\eqn\aag{
\widetilde \KK=- \frac{\VV_{10}}{2}\, \frac{R}{2(2\pi)^9} \int_0^\infty dl~
\frac{1}{\eta^8\left(il \right)}
\sum_{w\in 2\IZ+1} e^{-\pi l \frac{w^2R^2}{4}}
\sum_{a \in \IZ_2}(-)^a \frac{\vartheta^4\big[ {1 \atop a} \big]\left(il\right)}
{\eta^4\left(il \right)}
~.}
In this channel only odd winding modes in the RR sector contribute.
Since all of these modes are massive, we conclude that the O$'$9-plane
presented here has no massless tree level tadpoles. The non-tachyonic, 
type 0$'$B theories that will be presented below, share many similar features 
with this ten-dimensional example.

We can gain intuition about the geometry of this O$'$9 orientifold by looking at a similar
example in the conformal field theory of a single $U(1)$ boson $X$
with radius $R$. In that context, one can formulate four basic parities:
the usual worldsheet parity $\Omega$, the parity $s\Omega$, where
$s$ is again given by \aad, and the parities ${\cal I}\Omega$ and 
${\cal I}'\Omega$, which can be obtained by T-duality from the $\Omega$,
$s\Omega$ parities of the $2/R$ theory \BrunnerEM. Both ${\cal I}\Omega$
and ${\cal I}' \Omega$ include the involution $X\to -X$ and they have
two fixed points, one at $X=0$ and the other at $X=\pi R$. So, both parities 
involve a pair of O0-planes. In the ${\cal I}\Omega$ case, the orientifold
planes have the same tension, whereas in the ${\cal I}'\Omega$ case they have opposite
tension. In the case analyzed in this subsection, we are dealing essentially with
its T-dual, an $s\Omega$ parity, which, in the $S^1$ part of the theory,
we can think of as a combination of two circle-wrapping O1-planes with
opposite RR charge.

\newsec{Non-critical type 0$'$B strings}

Given the natural presence of 
dilaton tadpoles in non-supersymmetric vacua (at tree level and beyond),
it is, as explained in the introduction, a good idea to look for stable 
non-supersymmetric vacua with a non-trivial dilaton background. 
In this section, which contains the main results of this note, 
we will concentrate on backgrounds that possess asymptotically 
a linear dilaton. In general, these backgrounds have a weak coupling asymptotic region 
where the tree-level analysis is reliable (as the string coupling constant 
falls off exponentially there) and a strong coupling end 
where non-trivial physics takes place. In the precise setup that will be
analyzed below, the strong coupling region is effectively cutoff by the
addition of a non-trivial momentum condensate. The  
genus expansion is then controlled by an effective string coupling 
related to the value of the condensate. With appropriate tuning 
string theory in these vacua  becomes perturbative  and the higher
loop backreaction is accordingly suppressed everywhere and not just 
in the asymptotic region. 

Backgrounds with an asymptotic
linear dilaton direction are interesting for a number 
of reasons. It has been argued on general grounds \AharonyUB\
that string theory on these backgrounds is holographically dual
to a non-gravitational theory. A well known class of examples arises in type II
string theory, where linear dilaton backgrounds 
appear in the near-horizon geometry of NS5-brane
configurations \refs{\CallanAT, \GiveonPX}. 
The holographic dual in this case is a non-local,
non-gravitational theory known as Little String Theory \SeibergZK.

Another well known example is that of two dimensional non-critical 
strings. Type 0 strings in two dimensions have been discussed from the worldsheet and
matrix model point of view in \refs{\TakayanagiSM,\DouglasUP}.
In this case string theory is both perturbatively
and non-perturbatively stable. Orientifolds in this context have
been discussed in \refs{\GomisVI,\BergmanYP}. For  
the  0B$/\Omega$ orientifold, one finds 
no massless tadpoles from the Klein bottle.  For all other parities, 
there is a tachyon tadpole.\foot{The tachyon
in two dimensions, which is the only physical propagating
degree of freedom, is a massless scalar field.} 
Interestingly enough, it was pointed out in \GomisVI\
that the dual matrix model appears to describe string 
propagation in a shifted background, where the string 
divergences have been cancelled via the Fischler-Susskind
mechanism. From the  worldsheet point of view 
one  could attempt to cancel the tachyon tadpoles 
explicitly with the addition of the appropriate
number of D1-branes \BergmanYP. The D1-branes cancel
the massless one-loop divergence in the leading volume diverging
piece of the one-loop amplitude, but leave a finite uncanceled 
leftover \NakayamaEP. It was pointed out in \refs{\GomisVI,\BergmanYP}
that there is no type $0'$B $\hat c=1$ string in two
dimensions,
%one cannot project the type 0 $\hat c=1$ string in two
%dimensions with the parity $\Omega'=(-)^{\bar F}\Omega$ to obtain 
%a 0$'$B theory, 
because the $\NN=1$ Liouville potential is odd 
under $(-)^{\bar F} \Omega$. We will see that this obstruction can be evaded 
in a set of higher dimensional examples which we now proceed
to discuss.

In what follows, we will examine the possibility of stable type 0$'$
vacua in non-critical string theories in higher dimensions. 
In that case the NSNS ground state is really a tachyon that needs 
to be projected out in order to define a perturbatively stable 
theory. In this note we will focus on a class of non-critical string theories
that were first considered in \KutasovUA\ (for a review see
\refs{\KutasovPV}). These are string theories on
\eqn\aba{
\IR^{d-1,1}\times \left[ (\NN=2~ {\rm Liouville}) \times \MM\right]/\Gamma
~,}
where $d\leq 8$ is an even positive integer, 
$\Gamma$ is some discrete group and $\MM$ is the compact 
target space of some two-dimensional CFT with $\NN=(2,2)$
worldsheet supersymmetry ($e.g.$ a product of supersymmetric minimal models
or Landau Ginzburg models). 

The common factor of these backgrounds is $\NN=2$ Liouville theory. On the worldsheet
the latter is a theory of two interacting bosons $r,\varphi$ and two fermions
$\psi,\psi^\dagger$.\foot{In what follows daggers $^\dagger$ will be used
to denote complex conjugation in spacetime and bars right-moving
quantities on the worldsheet.} The $\NN=2$ Liouville action in superspace
notation (and $\alpha'=2$ conventions) reads
\eqn\abb{
\SS=\frac{1}{8\pi}\int d^2z\, d^4 \theta~ \Phi \Phi^\dagger+
\frac{\mu}{2\pi}\int d^2 z\, d\theta d\bar \theta ~ e^{-\sqrt{\frac{k}{2}}\Phi}
+\frac{\mu^\dagger}{2\pi} \int d^2 z\, d\theta^\dagger d\bar \theta^\dagger~
e^{-\sqrt{\frac{k}{2}}\Phi^\dagger}
~.}
In this expression $\Phi$ is the chiral $\NN=2$ superfield
\eqn\abc{
\Phi=r+i\varphi+i\sqrt 2 \theta \psi+i\sqrt 2 \bar \theta \bar \psi+
2 \theta \bar \theta F+ \cdots
~.}
The boson $r$ parametrizes a linear dilaton direction with linear dilaton
slope $\QQ=\sqrt{\frac{2}{k}}$. The central charge of the $\NN=2$ Liouville
theory is $c=3(1+\QQ^2)$, so the real parameter $k$ that appears 
in $\QQ$ and \abb\ is fixed by the criticality condition $c_{tot}=15$, where
$c_{tot}$ is the total central charge of the worldsheet theory \aba. 
The  compact $U(1)$ boson  $\varphi$ has the possible radii 
$R = n \QQ$, $n=1,2,\ldots$, such that the superpotential has the right periodicity properties.

By construction the background \aba\ exhibits $\NN=(2,2)$ worldsheet 
supersymmetry. At the special radius $R=\QQ$  one can construct spacetime 
supercharges which are mutually local
with all the vertex operators of a chirally GSO-projected theory. This chiral
GSO projection gives rise to a supersymmetric type II string theory on 
\aba\ with at least $2^{\frac{d}{2}+1}$ spacetime supercharges.
It is precisely this theory that appears as an equivalent description
in the near-horizon geometry of fivebrane configurations in ten-dimensional
type II string theory \GiveonZM.  

One can also impose a non-chiral GSO projection that gives rise to a
type 0 string theory on \aba. As their ten-dimensional 
non-supersymmetric cousins these theories
possess a closed string tachyon, hence they are perturbatively unstable.
Our objective here is to find an orientifold that projects out the tachyon
and leads to a type 0$'$ theory. We will focus on 
two specific examples with trivial $\MM$ and $\Gamma$: one at $d=4$ and
another one at $d=6$. The possibility of generalizations will be discussed
in the concluding section. The non-trivial part of the exercise is to
identify the appropriate orientifold in $\NN=2$ Liouville theory
and to analyze its properties. Orientifolds in $\NN=2$ Liouville
theory have been constructed recently in \IsraelSI.

Concluding this short introduction we would like to point out that
stable non-supersymmetric vacua can be constructed in theories
of the type \aba\ with appropriate deformations of type II string theory
on \aba. One such deformation arises when one changes the asymptotic
radius of $\varphi$. Usually in this case a closed string tachyon appears
infinitesimally close to the supersymmetric point, although there are some
interesting exceptions \ItzhakiZR. If some of the flat $\IR^{d-1,1}$ directions are 
compact, then the asymptotic background at infinity involves a 
higher dimensional torus. There is a certain subset of K\"ahler/complex
structure deformations of this torus that breaks spacetime supersymmetry,
but does not generate any closed string tachyons \HarmarkSF.

\subsec{A few words on  parities in  $\NN=2$ Liouville theory}

Before delving into specific examples, we would like to prepare the
ground with a few words on certain parities of the $\NN=2$ Liouville theory.
A general analysis of the orientifolds of the $\NN=2$ Liouville theory 
and its mirror dual supersymmetric $SL(2,\IR)/U(1)$ CFT can be found in \IsraelSI.

In a two dimensional QFT with $\NN=(2,2)$ supersymmetry there 
are two basic types of parities that reverse the worldsheet chirality
(exchanging the left- and right-movers) while preserving the holomorphy
of the $\NN=2$ supersymmetry. They are known as A- and B-type and 
the details of their definition can be found in \refs{\BrunnerZM,\IsraelSI}.
A distinguishing feature of these parities in the case of the $\NN=2$ 
Liouville theory is the following. A-type parities give orientifolds 
that are localized in the direction $\varphi$, whereas
B-type parities give orientifolds that are extended along $\varphi$. 
In the nomenclature of \IsraelSI\ the basic worldsheet parities 
are $\Omega_A$ and $\Omega_B$. 
$\Omega_A$ acts on the $\NN=2$ Liouville worldsheet fields as
\eqn\abca{\Omega_A \ : \ \
\Phi(z,\bar z) \to \Phi^\dagger(\bar z,z)~, ~ ~ 
\psi \to \bar \psi^\dagger~, ~ ~ \bar \psi \to \psi^\dagger
}
and $\Omega_B$ as
\eqn\abcb{\Omega_B \ : \ \
\Phi(z,\bar z) \to \Phi(\bar z,z)~, ~ ~ 
\psi \to \bar \psi~, ~ ~ \bar \psi \to \psi
~.}
The B-parity is related to the standard worldheet parity
$\Omega$, that appears in the previous section,
by $\Omega_B=(-)^{\bar F} \Omega$.

We will be interested in parities that project out the 
tachyonic NSNS ground state, but retain the $\NN=2$
Liouville interaction that appears in \abb.
The A-type parity $\Omega_A$ exchanges the two superpotential
terms in \abb\ and projects $\mu$ onto real values. However,
for the purposes of this work only  B-type orientifolds of $\NN=2$
Liouville  will be relevant. As explained in \IsraelSI, with B-type orientifolds
there are two obvious candidates for the construction
of type 0$'$B vacua. These are based on the $\NN=2$ Liouville
parities
\eqn\abcd{
\PP=s_\varphi \Omega_B~, ~ ~ 
\widetilde \PP=e^{\pi i \bar F}e^{-i\pi \bar J}\Omega_B
~,}
where $s_\varphi$ is the shift  \aad\ acting on $\varphi$ and
$\bar J$ is the right-moving $U(1)_R$ current. In the asymptotic
weak coupling region, where the $U(1)_R$ currents $J, \bar J$ take 
the simple form
\eqn\abi{
J=F+i\sqrt 2 \d \varphi~, ~ ~ \bar J=\bar F+i \sqrt 2 \bar \d \varphi
~,}
we can rewrite $\widetilde \PP$ as
\eqn\abj{
\widetilde  \PP=e^{-i\sqrt 2 \pi \bar p_{\varphi}} \Omega_B
~.}
where $\bar p_\varphi$ is the right-moving $\varphi$-momentum. 
Both $\PP$ and $\widetilde \PP$ project out the NSNS ground 
state, commute with the $\NN=2$ Liouville interaction and give
space-filling orientifolds. In the ensuing, we will focus on the former
parity which is geometric and has a clearer spacetime interpretation.

\subsec{Type $0'$B strings in four dimensions}
In this subsection we consider a special case of \aba\ with
$d=4$, $\MM$ trivial and $\Gamma$ the identity. In other words,
we want to consider type 0B string theory on 
\eqn\abd{
\IR^{3,1} \times \NN=2 ~{\rm Liouville}
~.}
In this case, $\NN=2$ Liouville theory has a background charge 
$\QQ=\sqrt 2$, {\it i.e.} $k=1$ in \abb. Extrapolating the arguments of \GiveonZM\
to type 0 string theory, it is natural to conjecture that type 0B string theory
on \abd\ provides in a double scaling limit a holographic description of the
Little String Theory that lives on two orthogonal NS5-branes in ten-dimensional
type 0B string theory. However, as the background is unstable the meaning
of this correspondence is unclear. 

The type 0B torus partition function of string theory on \abd\ can be found
in \refs{\MurthyES,\IsraelIR, \FotopoulosCN}. Here we will be interested only in the leading 
piece of the torus partition function associated to the weakly coupled
asymptotic linear dilaton region. It receives
contributions only from the continuous representations of $\NN=2$ Liouville
theory and is proportional to the infinite volume of the target space.
In terms of the continuous representation characters, defined for generic $k$ as
\eqn\abe{
ch_c\left(p,m;\tau\right)\left[ {a \atop b}\right]=
q^{\frac{p^2+m^2}{k}} \frac{\vartheta \left[ {a \atop b}\right](\tau)}{\eta^3(\tau)}\ ,
~}
the type 0B torus partition function on \abd\ reads
\eqn\abf{\eqalign{
\TT = & \frac{\VV}{2}  \int_\FF \frac{d\tau d\bar \tau}{4\tau_2}
\sum_{a,b \in \IZ_2} \frac{\vartheta\left[ {a \atop b} \right](\tau) 
\vartheta \left[ {a\atop b} \right] (\bar \tau)} 
{(8\pi^2 \tau_2)^2 \eta^3(\tau) \eta^3(\bar \tau)}
\ \times \cr 
& \times \ 
\sum_{n,w\in \IZ} \int_0^\infty dp\ 
ch_c\left( p, \frac{n+w}{2};\tau\right)\left[ {a \atop b}\right]
ch_c\left( p, \frac{n-w}{2};\bar \tau\right)\left[ {a \atop b}\right]
~.}}
The quantum numbers $n$ and $w$ represent the momentum and
winding around the angular coordinate $\varphi$ of $\NN=2$ Liouville
theory (the former is broken by the $\NN =2$ Liouville interaction in 
the strong coupling region). 
The lowest level spectrum of this theory can be found in table~1 of 
appendix~A of \MurthyES. It comprises of the graviton multiplet $(G,B,\phi)$,
a tachyon $T$, two non-negative mass squared RR scalars $C_0,C'_0$ and
a two-form potential $C_2$. All these modes appear with any momentum 
and winding $(n,w)$.

The worldsheet parity $\PP$, eqn.\ \abcd, projects out the tachyonic  
NSNS ground state with $(n,w)=(0,0)$, but keeps all the combinations of  
tachyon zero-modes of the form $|(n,w)\rangle -(-)^{n} |(n,-w)\rangle$, 
which are all massless or massive. 
In particular, for $w=0$, $\PP$ keeps all the modes with odd momentum $n$.
This is an appealing feature, since it allows for the $\NN=2$ Liouville 
potential --~with $(n,w)=(\pm 1,0)$~-- 
to be invariant. Hence, the parity $\PP$ is well defined in 
$\NN=2$ Liouville theory beyond the asymptotic region. Furthermore,
it was pointed out in \IsraelSI\ that under $\PP$ the cigar interaction
\eqn\abk{
\delta \SS=\mu_{cigar}\int d^2z \, d^4\theta ~e^{-\frac{\Phi+\Phi^\dagger}{\sqrt 2}}
}
is invariant. This property is required for the non-perturbative consistency
of the theory. 

The asymptotic Klein bottle amplitude
associated with $[\IR^{3,1} \times \NN=2 ~{\rm Liouville}]/\PP$ reads
\eqn\abl{
\KK=\frac{\VV}{2}\int_0^\infty \frac{dt}{2t} \sum_{a=0,1} (-)^a\sum_{n\in \IZ} \int_0^\infty dp~
(-)^{n+1} ch_c\left(p,\frac{n}{2}\right)\left[ {a \atop 1} \right] (2it)
\frac{\vartheta\left[ {a \atop 1}\right](2it)}
{(8\pi^2 t)^2 \eta^3(2it)}
~.}
Only zero winding states in the 
$\widetilde {\rm NS}$$\widetilde {\rm NS}$ and $\widetilde{\rm R}$$\widetilde{\rm R}$
sectors contribute to the trace. 
Adding the contributions from \abf\ and \abl\ we obtain, as advertised, 
a tachyon-free spectrum. 

In general, $\NN =2$ Liouville theory contains not only continuous 
representations, but also discrete representations describing 
states localized near the strong coupling
region of the theory. One may wonder whether 
some localized tachyon may show up in these sectors. 
Fortunately, the discrete representations appear in the range 
\eqn\ablb{\frac{1}{2} < j < \frac{k+1}{2}\, ,} 
and satisfy the conditions 
$p_\varphi/\QQ- j \in \IZ$ and  $\bar p_\varphi/\QQ- j \in \IZ$.
Therefore none of them appears in the model at hand.

In spacetime, the worldsheet parity $\PP$ gives rise to a space-filling 
O$'$5-plane. Hence, it is important to check if this orientifold has any
massless tadpoles. Using the well-known S-modular transformation properties of the 
continuous characters and the theta-functions (see $e.g.$ \IsraelSI)
we obtain from \abl\ the Klein bottle amplitude in the crosscap channel
\eqn\abm{
\widetilde \KK=-\frac{\VV}{8(2\pi)^4} \int_0^\infty dl\  
\sum_{a \in \IZ_2}
\frac{\vartheta\left[ {1 \atop a} \right]\left(il \right)}
{\eta^3\left(il \right)}
\sum_{w \in 2\IZ+1}  ch_c \left(0,\frac{w}{2};il\right)\left[{1 \atop a}\right]
~.}
This result is similar to the one obtained for our $\IR^{8,1} \times S^1$  model 
in critical string theory, eqn.\ \aag. Most notably, the orientifold sources again only odd winding 
modes in the RR sector. All these modes are massive and hence the 
O$'$5-plane in question does not exhibit any massless tree-level tadpoles. 
Consequently, this is an example of a non-supersymmetric theory 
of unoriented bosonic closed strings which is fully stable at tree-level as 
it exhibits neither tachyons nor tadpoles. 

In this subsection we did not construct the exact crosscap state 
associated with the parity $\PP$ in the background \abd, but this can be done
with the technology of \IsraelSI. In that paper, the 
crosscap state of a related A-type  orientifold  in  the mirror $SL(2,\IR)/U(1)$ was 
derived. We expect similar qualitative 
results for the O$'$5-plane presented here.
For more details see the closely related exact crosscap state
that appears at the end of the following subsection.

\vskip0.4cm
\noindent
{\it Comments on D-branes}
\vskip0.2cm

D-branes in type IIB string theory on $\IR^{3,1}\times SL(2,\IR)/U(1)$, made of 
B-type branes of $SL(2,\IR)/U(1)$ (equivalently A-type branes in type IIA string theory on \abd)
were used in \refs{\FotopoulosCN,\AshokPY,\MurthyXT} to engineer the electric 
and magnetic descriptions of $\NN=1$ SQCD with $N_c$ colours and $N_f$ flavors.
Metastable non-supersymmetric configurations were discussed
in this context in \MurthyQM. D-brane configurations with orientifolds were analyzed 
in \AshokSF. 

In this paper we consider a different theory, type 0B on $\IR^{3,1}\times \NN=2$ Liouville  
in the presence of an O$'$5 orientifold plane. The spectrum of
allowed D-branes in this theory contains the following possibilities
(see \HosomichiPH\ and references therein for a comprehensive analysis
of D-branes in $\NN=2$ Liouville theory):
\item{($i$)} D-branes that are localized in the strong coupling region
of the $\NN=2$ Liouville throat. These branes have A-type boundary
conditions in the $\NN=2$ Liouville part of the worldsheet CFT and
depending on the boundary conditions that are imposed on the extra four flat directions
we may have unstable D$p$-branes for $p=1,3$, where $p$ refers to the number of space-like dimensions filled by the brane in $\IR^{3,1}$,
or dyonic D$p$-branes with $p=0,2$.\foot{The word dyonic
refers here to the standard pair of an electric and a magnetic boundary state of
the type 0B theory. These branes do not have an open string tachyon on their worldvolume.
Notice that because of the A-type boundary conditions
in $\NN=2$ Liouville, one can have stable even-dimensional  
localized  D-branes in the non-critical type 0B theory, unlike in ten-dimensions.}
The latter exhibit a vanishing annulus amplitude.
All the branes have vanishing M\"obius strip amplitude with
the O$'$5-plane. Indeed, notice that these
branes, as A-type boundary states in $\NN=2$ Liouville,
are coherent states of momentum modes,
whereas the O$'$5-plane, as a B-type crosscap state, is a coherent state
of odd winding modes.
\item{($ii$)} D-branes that are extended in the linear dilaton direction
and are A-type in the $\NN=2$ Liouville throat. In this case, we can have
unstable D$p$-branes for any even $p\leq 4$, or dyonic D$p$-branes with odd
$p\leq 5$. As in case ($i$) all the branes here have a vanishing M\"obius
strip amplitude with the O$'$5-plane.
\item{($iii$)} D-branes that are extended in the linear dilaton direction
and are B-type in the $\NN=2$ Liouville throat. In this case, there are
unstable  D$p$-branes for any even $p\leq 4$ and dyonic D$p$-branes with odd
$p\leq 5$. These branes have a non-vanishing M\"obius strip
amplitude.

To illustrate the salient features of the above list we consider a few 
representative examples. Let us denote the boundary states that describe the 
D$p$-branes in case $(i)$ as $|Dp\rangle_{NS}$, $|Dp\rangle_{R}$ 
(in the NSNS and RR sectors respectively). Using the notation of 
\FotopoulosCN\ for the $\NN=2$ Liouville part of the boundary states
and the standard notation of \GaberdielJR\ for the fermionic part, 
we can write $|Dp\rangle_{NS}$ in terms of (extended) Ishibashi states 
as\foot{We quickly remind the reader that the fermionic boundary states
$|\eta\rangle_{NSNS/RR}$ exhibit the overlaps 
$_{NSNS}\langle \eta|\eta'\rangle_{NSNS} \sim \left[ 0 \atop \frac{1-\eta \eta'}{2} \right]$,
$_{RR}\langle \eta|\eta'\rangle_{RR} \sim \left[ 1 \atop \frac{1-\eta \eta'}{2} \right]$,
where the rhs denotes the sector $\left[ a \atop b \right]$ in which the standard
fermionic characters appear.}
\eqn\baa{\eqalign{
|Dp\rangle_{NS}=&\int_0^\infty dP~ \Big[ \Phi_{NS}(P,0;+)
\big( |P,0;+\rrangle_{NSNS}-|P,0;-\rrangle_{NSNS} \big)+
\cr
&+\Phi_{NS}\left(P,\frac{1}{2};-\right) 
\big( |P,\frac{1}{2};+ \rrangle_{NSNS}+|P,\frac{1}{2};-\rrangle_{NSNS}\big)\Big]
~.}}
Using the known action of $\Omega$, $(-)^{\bar F}$ and $s_{\varphi}$
on the boundary states \refs{\GaberdielJR,\BergmanRF}
\eqn\bab{
\Omega |P,m;\eta\rrangle_{NSNS}=|P,m;\eta\rrangle_{NSNS}~, ~ ~ 
\Omega |P,m;\eta\rrangle_{RR}=-(i\eta)^{2-p} |P,m;\eta\rrangle_{RR}
~,}
\eqn\bac{
(-)^{\bar F}|P,m;\eta\rrangle_{NSNS}=-|P,m;-\eta\rrangle_{NSNS}~, ~ ~
(-)^{\bar F}|P,m;\eta\rrangle_{RR}=|P,m;-\eta\rrangle_{RR}
~,}
\eqn\bad{
s_\varphi |P,m;\eta\rrangle_{NSNS/RR}=(-)^{2m} |P,m;\eta\rrangle_{NSNS/RR}~,  ~ ~ 
m=0,\frac{1}{2}~, ~ ~ \eta=\pm
~}
we deduce that $|Dp\rangle_{NS}$ is invariant under 
$\PP=s_{\varphi}(-)^{\bar F}\Omega$.
This boundary state describes for odd $p$ an unstable brane,
which includes an open string tachyon in its spectrum. For even $p$
this NSNS boundary state alone describes a D$p$-$\bar {\rm D}p$ pair as in ten dimensions.
The wavefunctions $\Phi_{NS}(P,0;+)$, $\Phi_{NS}(P,\frac{1}{2};-)$ appear in a
type II context in \FotopoulosCN.

To obtain branes without open string tachyons one needs to add a RR sector contribution
to the NSNS sector D-brane boundary state.  This is possible
only for even $p$ as for odd $p$ the resulting boundary state is not GSO invariant.
This is an unfortunate feature 
for the construction of interesting four dimensional gauge theories in 
this non-supersymmetric setup 
(such theories would be realized on D3-branes at the strong coupling
region of the $\NN=2$ Liouville throat). It should have been expected, however, since
we are considering a type 0B  theory and the only known
branes with ``Dirichlet'' boundary conditions in the linear dilaton direction
are the ones that are A-type in $\NN=2$ Liouville theory.

For even $p$ we consider the generic  linear superposition
of Ishibashi states in the RR sector:\foot{See \FotopoulosCN\ for the exact expressions
of the R sector wavefunctions $\Phi_R(P,0;+)$, $\Phi_R(P,\frac{1}{2};-)$.}
\eqn\bae{\eqalign{
|Dp\rangle_R=&\int_0^\infty dP~ \Big[ \Phi_R(P,0;+) 
\big(  |P,0;+\rrangle_{RR}+\alpha |P,0;-\rrangle_{RR} \big)+
\cr
&\qquad +\Phi_R\left(P,\frac{1}{2};-\right) 
\big(  |P,\frac{1}{2};+\rrangle_{RR}+\beta |P,\frac{1}{2};-\rrangle_{RR}\big) \Big]\ .
}}
This RR boundary state is invariant under the parity  $\PP$  iff
\eqn\baf{
\alpha=i^{-p}  ~, ~ ~ \beta=-i^{-p}
~.}
So we find for $p=0,2$ the $\PP$-invariant boundary states 
$|Dp\rangle_{NS}\pm |Dp\rangle_R$
(the upper and lower sign refer to D-branes and anti-D-branes respectively). They are identical to 
the boundary states of the corresponding dyonic D$p$-branes in the type 0B theory 
on \abd\ and give a vanishing annulus amplitude. 
We find that a stack of $N$ such D-branes  realize a
non-supersymmetric gauge theory with $U(N)$ gauge group,
an adjoint real scalar, as well as 
fermions in the symmetric and antisymmetric representations. This theory is a non-supersymmetric
descendant of the three-dimensional $\NN=2$ SYM; the latter
 can be realized in  type IIB string theory on \abd\ with the corresponding 
BPS D2-branes.\foot{Flavors can be added to both of these theories 
with the addition of the appropriate D4-branes to obtain three dimensional
$\NN=2$ SQCD or its non-supersymmetric descendant.
It would be interesting to consider Seiberg duality for these gauge theories
(especially the non-supersymmetric one)
in the setup of the non-critical string generalizing the arguments of
\refs{\MurthyXT,\AshokSF}.}

The above exercise can be repeated {\it mutatis mutandis} for A-type D-branes of 
$\NN=2$ Liouville theory that are extended in the linear dilaton direction. 
With a few modifications similar results can also be obtained for extended 
D-branes that are  B-type in  $\NN=2$ Liouville theory (boundary states for such branes can be 
deduced easily from the boundary CFT analysis of \HosomichiPH). 
In this case, there is a non-vanishing M\"obius strip amplitude
that can be determined with the use of the exact crosscap wavefunction 
that appears in the following subsection (appropriately adapted to the 
four dimensional case). We will not, however, pursue this calculation
any further here.

\subsec{Type $0'$B strings in six dimensions}

We now turn to an example with
six dimensional Poincar\'e invariance that involves type 0B string theory 
on
\eqn\aboaa{
\IR^{5,1}\times \NN=2 {\rm \ Liouville}.}
It describes the near-horizon limit of a pair of parallel fivebranes.
This is a special case of the CHS background $\IR^{5,1} \times \IR_\QQ \times SU(2)_k$
(the near-horizon geometry of $k$ parallel NS5-branes  \CallanAT),
with $k=2$. At this particular value of the level $k$,
the supersymmetric $SU(2)$ WZW model contains only an
$SU(2)_2 \times SU(2)_2$ algebra consisting of three worldsheet fermions
(both left- and right-handed), as the bosonic WZW model becomes trivial.
After the bosonization of the two complex fermions
$\psi^{\pm}$ and $\bar{\psi}^{\pm}$ (the superpartners
of the $SU(2)$ currents $J^\pm$ and $\bar{J}^{\pm}$) we obtain a $U(1)$
affine algebra at level $k=2$. This observation 
allows us to rewrite the torus amplitude for the type 0B theory as:
\eqn\abna{\eqalign{
\TT =& \frac{\VV}{2} \int \frac{d \tau d \bar \tau}{4\tau_2} \frac{1}{(8\pi^2 \tau_2)^3}\!\!
\sum_{a,b \in \IZ_2}\!\! \frac{|\vartheta \left[ {a \atop b}\right]|^4}
{|\eta|^{12}} \ \times \cr & \times \ \int_0^\infty \!\!\! dp 
\sum_{r, \bar r \in \IZ} (-)^{b(r-\bar r)} ch_c  (p,r+\frac{a}{2};\tau)\left[ {a \atop b}\right] 
\bar{ch}_c (p,\bar r+\frac{a}{2};\bar \tau)\left[ {a \atop b}\right] \, .  
}}
We identify the momentum $n$ and the winding number $w$ in the $\NN=2$ Liouville as
\eqn\aboa{
n = \frac{r+\bar r+a}{2}  \ \  , \ \ \  w = r - \bar r
~.}
Fractional momenta are allowed since the generalized diagonal projection
acts as a $\IZ_2$ winding shift orbifold of the
transverse direction $\varphi$ in $\NN=2 {\rm \ Liouville}$.\foot{It 
should be noted that there is another possible type 0B  modular invariant,
where the $\IZ_2$ orbifold acts only on the fermions and not on the
$U(1)_2$ boson $\varphi$.  Both theories are consistent (both exhibit mutual locality of operators),
but only the first one describes 
the near horizon geometry of two parallel fivebranes \MurthyES. 
In what follows, we will consider only this case.}

As the torus partition function \abna\ is very closely related to that of the 
critical ten-dimensional type 0B superstring, one can write down a Klein bottle 
amplitude for the 0$'$B theory which is similar to that of \refs{\BianchiYU,\SagnottiGA,\SagnottiQJ}
(modulo two $\eta$-functions). In terms of the $\NN=2$ Liouville continuous characters \abe\
we obtain
\eqn\abqa{
\KK=-\frac{\VV}{2} \int_0^\infty \frac{dt}{2t}~
\frac{1}{(8\pi^2 t)^{3}}  \sum_{a \in \IZ_2} (-)^a
\frac{\vartheta^2\big[ {a \atop 1} \big](2it)}{\eta^6(2it)}\int_0^\infty 
\!\!\! dp  \, \sum_{n \in \IZ}
e^{i\pi (n+a/2)}
ch_c (p, n+a/2;2it) \left[ {a \atop 1} \right]\, .}
Alternatively, one can derive this amplitude from the worldsheet parity 
$\PP$ in \abcd. This justifies {\it a posteriori} the choice of a B-parity
of $\NN=2$ Liouville in order to construct a type 0$'$ string theory in the non-critical
background \aboaa .
As the theory involves half-integral momentum one may wonder
whether this parity is involutive. However, the potentially non-trivial phases
occur in the RR sector ($i.e.$ $a=1$), where the fermions contribute
to $\PP^2$ as $\exp\, 3i\pi/2$, hence we deduce that the
parity is indeed involutive.

Adding the torus and Klein bottle amplitudes \abna\ and \abqa\
respectively, we obtain as anticipated a
tachyon-free continuous spectrum.\foot{It is  interesting to remark that, without
the non-trivial action of the GSO projection on the compact boson $\varphi$  of $\NN=2$ Liouville
(in the type 0B model we consider),
one would have winding tachyons surviving the orientifold projection.}
As before there may be localized tachyons in the discrete representations.
Using \ablb\ one finds that the only physical discrete states have $j=1$. The lowest
dimension operator is the $\NN=2$ Liouville chiral primary
with $n=1$ and $w=0$, of dimension $L_0 = j/k=1/2$. Hence it gives
a massless state in spacetime, which is none other than
the momentum condensate in the $\NN=2$ Liouville action, eqn.\ \abb.

The six dimensional type 0$'$B theory is also free of massless tadpoles.  
In the transverse channel, the Klein bottle amplitude \abqa\ reads:
\eqn\abta{
\widetilde{\KK} = - \frac{\VV}{8(2\pi )^{6}} \int_0^\infty dl~
\sum_{a \in \IZ_2}(-)^a
\frac{\vartheta^2\left[ {1 \atop a} \right](il)}{\eta^6(il)} \, \sum_{w \in 2\IZ+1 }
(-)^{\frac{a(w-1)}{2}}
ch_c \left( 0, \frac{w}{2};il \right) \left[ {1 \atop a} \right]
~}
coupling only to odd winding states in the RR sector. In the $l\to \infty$ limit,
the dominant contribution to this amplitude comes from states
with unit winding $w=\pm 1$ and $\widetilde \KK$ behaves like
\eqn\abua{
\tilde{\KK} \sim -\frac{\VV}{8(2\pi )^{6}} \int_0^\infty dl~ e^{-\frac{\pi l}{4}}
\left[ 1 + \OO ( e^{-2\pi l } ) \right]
~.}
We conclude that the RR tadpoles are again massive. As in the previous
four dimensional example, this is due to the "universal" mass shift
$m_{\rm min}^2 = \QQ^2/4$ in linear dilaton backgrounds 
(for the continuous representations).\foot{This 
property of linear dilaton backgrounds was also used in \KiritsisTA\ 
as an infrared regulator in closed string one-loop computations.}
Hence, we obtain another example of a closed 
unoriented theory which is perturbatively stable,  tadpole-free
and does not require the addition of open string sectors for consistency.

One can construct various D-branes in the six-dimensional non-critical type 0$'$B
theory, along the lines of our analysis of the four-dimensional model. There is however
an important difference. Because of the non-trival GSO projection, and the enlarged
$\NN=(4,4)$ superconformal symmetry of $\NN=2$ Liouville theory with $\QQ=1$, 
it is actually possible to evade the constraints found in four dimensions and 
obtain stable D-branes filling $\IR^{3,1}$, localized in the linear dilaton direction.
This allows to engineer four-dimensional, non-supersymmetric gauge theories (with
fermions).  We postpone the detailed study of the properties of these 
theories to future work.

\vskip0.4cm
\noindent
{\it The exact crosscap state}
\vskip0.2cm

Crosscap states in $\NN=2$ Liouville theory and its mirror dual
$SL(2,\IR)/U(1)$ supersymmetric coset were constructed in \IsraelSI.
Using the results of that paper (more specifically the wavefunction for the 
B-type parity of $\NN=2$ Liouville),
we obtain the following candidate for the type 0$'$B crosscap wavefunction:
\eqn\abv{ \eqalign{
\Psi (p^\mu,P,w) \left[ b \atop a \right] 
\propto & \ e^{\frac{i\pi}{2} (1-a)+\frac{i\pi a(w-1)}{4}}\,
\delta_{b,1 \mod 2} \, \delta_{w,1 \mod 2}\, 
\delta^6 (p^\mu) \ \times \cr & \times \ 2^{-iP} \cosh \frac{\pi P}{2}\
\frac{\Gamma (1-iP)\, \Gamma(-2iP)}{\Gamma (
\frac{1}{2} - iP + \frac{w-1}{2})\, \Gamma ( \frac{1}{2} - iP - \frac{w-1}{2})},}}
up to a normalization constant that is fixed by channel duality.
The labels of the wavefunction $\Psi$ on the rhs of this equation
denote in an obvious fashion the quantum numbers of the closed
string vertex operators whose one-point function on the $\IR {\rm P}_2$
we are computing. A similar expression holds for the orientifold of the 
four dimensional non-critical string theory considered in the previous subsection.

\newsec{Discussion}

In this note we presented two tachyon-free examples of type 0$'$B closed
string theories without massless tree-level tadpoles on backgrounds with non-trivial
dilaton profiles. We would like to conclude with a list of interesting questions
and related comments.

\vskip0.4cm
\noindent
$(a)$ {\it Fivebranes,  double-scaling limits and LST's  in type 0}
\vskip0.2cm

Backgrounds of the form \aba\ can be embedded naturally
in ten-dimensional type II string theories as backgrounds that
describe string theory dynamics
in the near-horizon region of fivebrane configurations or in the vicinity
of Calabi-Yau singularities in a double scaling limit \GiveonZM. The simplest
example, that we discussed in the previous section, corresponds to  parallel fivebranes. 
As reviewed there,  the six dimensional non-critical string arises as a special, 
degenerate case within a family of $critical$ string backgrounds describing 
the near-horizon limit of $k$ parallel fivebranes.\foot{Similarly
type II string theory on $\IR^{3,1} \times \NN=2$ Liouville
describes string propagation in the near-horizon region of two orthogonal
NS5-branes, or in the vicinity of the conifold singularity.}
A natural question concerns the embedding of type 0$'$B 
non-critical string theories in ten dimensions with an
appropriate configuration of fivebranes and orientifolds.
Clearly, the ten-dimensional theory cannot be type 0$'$B, 
because in that theory the NSNS  two-form  is projected
out and the theory cannot accommodate any NS5-branes.\foot{It is worth noticing
that in the non-critical type 0$'$B theories of the previous section certain 
modes of the NSNS two-form are not projected out and remain
in the physical spectrum.}

An important clue comes 
from the classification of orientifolds in the $SU(2)$ WZW model,
see {\it e.g.} \BachasID. Consistent orientifolds in this theory by definition
preserve the Wess-Zumino term and can be
either a pair of antipodal O0-planes or  an equatorial O2-plane. In order
to generalize the six-dimensional non-critical type 0$'$B theory to arbitrary
values of $k$, one can build a type 0A model on the general CHS
background $\IR^{5,1}\times \IR_{\QQ} \times SU(2)_k$ with an
orientifold that is space-filling in $\IR^{5,1}\times \IR_{\QQ}$ and has the geometry of
a pair of O0-planes in $SU(2)_k$. This setup is such that it
can be interpreted as the near-horizon limit of $k$ NS5-branes
at the origin of the transverse space $x^{6,7,8,9}$ in the unoriented string theory
${\rm 0A}/({\cal I}_{678}(-)^{\bar F}\Omega)$.\foot{A type IIA
orientifold of the CHS background with the same geometry
was studied in \refs{\ForsteUR,\BurshtynXZ}.} Indeed, in spherical coordinates
for the transverse space, the orientifold is extended along the radial (linear dilaton)
direction and intersects the three-sphere at two antipodal points.  

For $k$ even one can spread the fivebranes evenly on a circle in the 
$(x^6,x^7)$ plane in the presence of the orientifold.\foot{For $k$ odd
the distribution of fivebranes is not symmetric under ${\cal I}_{678}$.}
This amounts to adding the $\NN=2$ Liouville interaction on the worldsheet
theory of $\IR_{\QQ} \times SU(2)_k$, see the last two terms in eqn.\ \abb.
By T-duality --~along the angular coordinate in the 
$(x^8,x^9)$ plane~-- one gets type 0B in $[
\, \NN=2\ {\rm Liouville}\, \times SU(2)_k/U(1)\, ]/\IZ_k$.  In this
mirror background the orientifold is made of the B-type O2-plane 
of $\NN=2$ Liouville \IsraelSI, and the point-like 
B-type O0 orientifold of $SU(2)/U(1)$ that exists for 
$k$ even \BrunnerZM. The Klein bottle amplitude for the resulting model reads:
\eqn\baa{\eqalign{
\KK = & -\frac{\VV }{2} \int \frac{dt}{2t} \sum_{a} (-)^a \frac{1}{(8\pi^2 t)^3}
\frac{\vartheta^2 \left[{a \atop 1}\right]}{\eta^6} 
\ \times \cr & \ \times \ \int_0^\infty \!\!\! dp \sum_{m \in \IZ_{2k},w \in \IZ}\!\!\!\!\!
ch_c \left(p,\frac{m}{2}+kw;2it\right)  \left[{a \atop 1}\right]
\sum_{2j=0}^{k-2} (-)^m C^{j}_{m}  \left[{a \atop 1}\right] (2it)
}}
where  $C^{j}_{m} [{a \atop b}]$ are characters of the super-coset $SU(2)/U(1)$. 
At $k=2$, the supercoset $SU(2)/U(1)$ contains only the identity. Hence,
the following relation holds
\eqn\bab{
\sum_{j}  C^{j}_{m}  \left[{a \atop 1}\right] = e^{-\frac{i\pi a}{2}} \left[ 
\delta_{m,a\, \mod \  4} -\delta_{m,a+2\, \mod \ 4}  \right]\  \ \ \ {\rm for}\ \ k=2
~.} 
Consequently, one finds that the Klein bottle amplitude in the background of two fivebranes,
eqn.\ \baa, becomes precisely the same as that of the type 0$'$B non-critical strings in
six dimensions, eqn.\ \abqa.\foot{Note that if one had chosen instead an O2-plane
of $SU(2)$ --~giving an O2-plane of $SU(2)/U(1)$~-- there would be an extra phase
$(-)^a$ in \abqa. This would reverse the orientifold projection
in the RR sector.} Notice that the (asymptotic) translation 
symmetry $\varphi \to \varphi + \lambda$  
in $\NN=2$ Liouville corresponds to rotations in the
$(x^{6},x^{7})$ plane where the two NS5-branes are separated,\foot{Only 
a $\IZ_2$ subgroup is preserved by the $\NN=2$ Liouville interaction
in the GSO-projected theory \abna, as the fivebranes
break the rotational symmetry.} while the winding symmetry
$\tilde \varphi \to \tilde \varphi + \tilde \lambda$
corresponds to rotations in the $(x^{8},x^{9})$ plane.
As the fixed point of $s_\varphi$ is the origin of the $(x^{6},x^{7})$ plane,
and the crosscap has couplings to winding modes,
this is in accordance with the conclusion that the geometry of
the orientifold is $x^{6,7,8}=0$,
$i.e.$ that it is an O$'$6-plane in the fivebrane geometry.
These observations can be used as further evidence in favor of
the crosscap appearing in eqn.~\abv. Moreover, one
can add in type 0A non-tachyonic D4-branes stretched between the two parallel 
NS5-branes, along $x^6$. In the presence of the O$'6$ plane, this configuration 
leads to a non-supersymmetric four-dimensional gauge theory
as mentioned in the previous section.

One may wonder whether generically type 0A string theory in the near-horizon 
geometry of $k$ fivebranes with an O$'$6 orientifold is also perturbatively stable.
In flat space, the orientifold $0{\rm A}/({\cal I}_{678}(-)^{\bar{F}}\Omega)$
contains a "bulk" tachyon corresponding roughly speaking to closed strings away
from the O$'$6-plane. In general, adding $k$ NS5-branes and going to the
near-horizon limit --~the CHS background~-- does not help.
A tachyon appears in the representation $j=1/2$ of $SU(2)_{k,L} \times SU(2)_{k,R}$ (as
$2j+m=0\,~ \mod\,  2$ in the NS sector).\foot{For $k>2$ there may be also
localized tachyons in the spectrum, $i.e.$ made with the discrete representations
of $\NN=2$ Liouville.}
At $k=2$ only the "$s$-wave" ({\it i.e.} $j=0$) remains in the physical spectrum and
therefore the tachyon is gone. This shows that the tachyon-free six dimensional type 0$'$
non-critical string on $\IR^{5,1}\times \NN=2$ Liouville
is an exceptional case within the class of backgrounds that appear
in the near-horizon geometry of $k$ parallel fivebranes.

The four-dimensional non-critical string is also expected to be related to 
a type 0 critical string theory in the background of two orthogonal fivebranes with some
O$'$-plane. However, as this model does not belong to a family of critical
backgrounds with a similar interpretation ($i.e.$ more than two orthogonal fivebranes)
there is no direct way to prove this statement.

One can use the non-critical type 0$'$ vacua of this note as
the holographic definition (in the sense of \refs{\AharonyUB,\GiveonZM})
of a new kind of ``bosonic'' Little String Theories.
It would be interesting to understand certain aspects of these dual theories,
for instance, their low-energy dynamics. It would be also nice to elaborate
further on other four-dimensional non-critical strings of the form \aba\ with non-trivial 
CFT $\MM$.\foot{With $SU(2)_k/U(1)$, giving the CHS background, 
we exhausted all the possibilities in six dimensions up to discrete
identifications.}

\vskip0.4cm
\noindent
$(b)$ {\it Comments on asymptotic supersymmetry and holography}
\vskip0.2cm

An interesting general question in string theory is the following: how
badly can supersymmetry be broken in a stable vacuum?
In string theory this question is important in the context of the general
problem of supersymmetry breaking and vacuum selection, the problem of the
cosmological constant and for holography in non-supersymmetric situations.
For instance, strongly coupled QCD in the limit of a large number of colors is believed to
be a stable theory with an exponential density of weakly interacting hadrons 
\refs{\WittenKH,\Coleman}. This theory is also expected on general grounds \tHooftJZ\
to have a dual formulation as a string theory
(although the precise string theory with this property is not presently known).
This example would suggest that there are stable weakly interacting vacua of string
theory where supersymmetry is altogether absent and the spectrum comprises only
of bosons. Indeed, a number of such theories were discussed in this note, although
we only presented examples with unoriented closed strings. As we now review,
there is a deep reason for this feature.

At weak coupling one can make a number of interesting general statements
concerning the spectrum of stable string theories.
At the level of the torus amplitude there can be only
one kind of divergence (in more than two dimensions), an IR divergence due to a closed
string tachyon. Then, under very mild assumptions,
one can argue on the basis of worldsheet modular invariance
\refs{\KutasovSV,\KutasovPV} that a tachyon divergence is directly related to
an exponential mismatch between the asymptotic high energy density of bosons
and fermions. When the tachyon is absent, the asymptotic high energy density
of bosons will in general balance the number of fermions up to a two-dimensional
field theoretic density of states. This asymptotic cancellation between the number
of bosons and fermions has been dubbed {\it asymptotic supersymmetry}
(or misaligned supersymmetry in \refs{\DienesES,\DienesPM}).
Hence, the theorem of \KutasovSV\ places a non-trivial constraint,
the constraint of asymptotic supersymmetry, on how badly non-supersymmetric 
a stable theory of oriented strings can be at weak coupling.\foot{Since 
string theory on RR backgrounds is much less understood, it is not immediately 
clear if this theorem can be evaded in this case.}
It should be emphasized that the criterion of asymptotic supersymmetry 
is a purely stringy effect that is not visible in the low energy supergravity 
description of the theory.

In this paper we discussed a number of perturbative string theory vacua,
where asymptotic supersymmetry is altogether absent, but the theory is nevertheless
stable at tree level. The extra ingredient that allowed us to evade the general
one-loop arguments of \KutasovSV\ is the contribution of the Klein bottle
to the closed string one-loop vacuum amplitude, which
in certain cases can project out the would-be closed string tachyon.
Hence, we may conclude that asymptotic supersymmetry in the bulk
is not, in general, a necessary condition for tree level stability in string theory.
Nevertheless, we would like to argue here that the stability of closed strings
requires asymptotic supersymmetry in a slightly different form.

Using general arguments based on worldsheet duality, one can argue for a connection
between asymptotic supersymmetry in the open string sectors associated
with the various D-branes accomodated by a given string theory and IR instabilities in the bulk
\NiarchosKW. More specifically, an open string theory without
asymptotic supersymmetry is necessarily coupled to a $closed$ string tachyon,
which one can argue cannot be decoupled even in the low energy limit.
It should be noted that open string tachyons do not partake in this connection
with the asymptotic high energy density of open string states, instead they signal
the existence of lower energy configurations for the D-branes in question.
We would like to propose that when a theory has a non-trivial spectrum of D-branes, 
asymptotic supersymmetry on the branes is a more fundamental property than asymptotic
supersymmetry in the bulk as far as the stability of the theory in the bulk is 
concerned.\foot{A similar phenomenon has been observed in matrix theory,
where asymptotic supersymmetry seems to be a crucial requirement for 
locality and cluster decomposition \BanksMN.}
A natural general conjecture for which there is no known 
counterexample is the following:\foot{VN would like to thank Adi Armoni
for a discussion on this point.}

\vskip0.2cm
\noindent
{\bf Conjecture:} For theories that do not admit open strings, like the heterotic
string, asymptotic supersymmetry in the bulk is necessary and sufficient 
for tree-level stability by the standard arguments of \KutasovSV. For theories
with oriented or unoriented strings in the bulk that admit open strings, the closed
string spectrum is tachyon free if and only if all the open string sectors
associated with the allowed D-branes of the theory exhibit
asymptotic supersymmetry.

\vskip0.2cm
\noindent
Strong evidence for the validity of this conjecture arises from the combined
arguments of \refs{\KutasovSV,\NiarchosKW}. 

The above observations can have indirect consequences for the engineering
of non-supersymmetric gauge theories and holography in string theory.
Suppose we want to engineer a gauge theory as the low energy limit of an open
string theory on a stack of D-branes in a background without closed string tachyons. 
Then, according to the above conjecture the open string theory on the D-branes
will necessarily exhibit asymptotic supersymmetry.\foot{Even in backgrounds with closed
string tachyons we should consider D-brane configurations with an asymptotically
supersymmetric open string spectrum. Otherwise it is impossible to decouple
 open strings from the closed string tachyon in the usual low-energy decoupling
limits~\NiarchosKW.} This is a property of the high-energy degrees of freedom of string theory,
so it is difficult to see precisely how this property will affect the low-energy field-theoretic
degrees of freedom. At least one can deduce with certainty that
the full open string theory spectrum on the branes will necessarily include both bosons
and fermions. In the usual fermionic string this would imply the presence of both
NS and R sectors. Hence, if we are looking for a setup that realizes a purely bosonic
gauge theory (for example four dimensional bosonic Yang-Mills theory), somehow we have
to find a way to give a (stringy) mass to the fermions in the R sectors and it is not
immediately clear if and how this could be achieved.\foot{Of course, one could consider
D-brane setups where the low energy theory is a gauge theory with both bosons
and fermions. In this context one can give an arbitrary bare mass to the fermions,
$e.g.$ by anti-periodic boundary conditions for the fermions on an extra compact
direction, and then flow to the IR, where presumably one obtains a purely 
bosonic theory. This is not what we have in mind here. 
What we are looking for are D-brane setups that realize a bosonic gauge theory
at all scales after sending $\alpha' \to 0$.} A few examples that verify this 
difficulty include the dyonic branes in type 0B string theory in ten dimensions and
the D-branes in type 0$'$ theory in ten dimensions or in non-critical dimensions
(see sections 2 and 3 above). It would appear that perturbative stability 
in closed string theory comes with a set of constraining conditions and not everything
goes as one might imagine from low energy effective action intuitions.
It would be very interesting to get a better handle on this set of constraining 
conditions (the above conjecture is an example of such constraints).

\vskip10mm
\centerline{\bf Acknowledgments}
\vskip4mm

We would like to thank Carlo Angelantonj, Adi Armoni, Emilian Dudas and
David Kutasov for useful discussions and comments.
VN acknowledges partial financial support by the EU under the contracts
MEXT-CT-2003-509661, MRTN-CT-2004-005104 and MRTN-CT-2004-503369.

\listrefs
\bye